\documentstyle[11pt,aaspp4]{article}
\input psfig
\psfull
\singlespace
\def\be     {\begin{equation}}
\def\ee     {\end{equation}}
\def\ba     {\begin{eqnarray}}
\def\ea     {\end{eqnarray}}
\DeclareRobustCommand{\wisk}[1]{\ifmmode{#1}\else{$#1$ }\fi}

\def\ol    {\wisk{{\Omega_{\Lambda}}}}
\def\om    {\wisk{\Omega_{m}}}

\def\h     {\wisk{h}}

\def\obh   {\wisk{\Omega_{b}h^{2}}}
\def\ob    {\wisk{\Omega_{b}}}

\def\n     {\wisk{n}}
\def\Q     {\wisk{Q_{10}}}

\def\lp    {\wisk{\ell_{peak}}}

\def\l     {\wisk{\Lambda}}

\def\etal  {{\sl et al.}~\rm}
\def\apj   {{\sl Ap.J.$\!$}~\rm}

\newcommand {\lsim}{\mbox{$\:\stackrel{<}{_{\sim}}\:$} }
\newcommand {\gsim}{\mbox{$\:\stackrel{>}{_{\sim}}\:$} }

\begin{document}
\title{The Cosmic Microwave Background and Observational Convergence in the $\om - \ol$ Plane}

\author{ Charles H. Lineweaver \altaffilmark{1}\\
School of Physics, UNSW, Sydney 2052, Australia}
\altaffiltext{1}{charley@bat.phys.unsw.edu.au}
\date{Received ;     accepted        }
\begin{abstract}

I use the most recent cosmic microwave background (CMB) anisotropy 
measurements to constrain the leading cold dark matter models in 
the $\om - \ol$ plane. A narrow triangular region is preferred.
This triangle is elongated in such a way that its intersection
with even conservative versions of recent supernovae, 
cluster mass-to-light ratios and double radio source constraints,
is small and provides the current best limits in the $\om -\ol$ plane: 
$\ol = 0.62 \pm 0.16$ and $\om = 0.24 \pm 0.10$. 
This complementarity between CMB and other observations rules out $\ol = 0$ 
models at more than the $99\%$ confidence level.

\end{abstract}

\section{INTRODUCTION}
\label{sec:introduction}

The main goal of CMB measurements and the two new satellite missions
MAP and Planck Surveyor is to determine a host of cosmological parameters at the 
unprecedented accuracy of a few percent (Jungmann \etal 1996,
Zaldarriaga \etal 1997, Bond \etal 1997).
As part of this goal it is important to keep track of what can already be determined 
from the CMB without conditioning on certain families of models or on certain values 
of parameters within these families.
In this {\it Letter}, the analysis of CMB anisotropy measurements is expanded to include 
the most popular families of cold dark matter (CDM) models such as flat, flat-$\Lambda$ and open models, 
as well as the less popular open$-\Lambda$ models.
Figure 1 presents an overview of this parameter space.
Open-\l models are considered here because they subsume all of the
above models and thus  provide a parameter space in which
the most popular models can be directly compared.

The popularity of non-zero $\ol$ models has waxed and waned over the years;
for excellent reviews see Felten \& Isaacman (1986) and Carroll \etal (1992).
$\ol$ was introduced by Einstein (1917) to solve the discrepancy between
an apparently static universe and the dynamic cosmology of general relativity.
Since this inauspicious beginning $\ol$ has been invoked many times and
seems to be a surprisingly multi-purpose cure-all for theory-observation mismatches.
Several recent papers (Turner 1991, Ostriker \& Steinhardt 1995, 
Roos \& Harun-or-Rachid  1998), have pointed out the  effectiveness of $\ol$ in
resolving apparent conflicts between various observational constraints.
Recently, $\ol$ has been invoked to solve the discrepancy between 
globular cluster ages and the age of the Universe inferred from measurements of
Hubble's constant.

Recent supernovae results in $\ol = 0$ models yield $\om$ values so low that they are
unphysical:  $\om = -0.4 \pm 0.5$ (see column 2 of Table 1, error bars and limits in this
{\it Letter} are $68\%$ confidence levels unless stated otherwise). 
Not only are they unphysically low but the highest $\om$ values allowed by the error
bars are in  strong disagreement with the high values of $\om$ preferred by
the CMB in these same models (Lineweaver 1998). In Lineweaver \& Barbosa (1998b) we
report a 99.9\% confidence lower limit of $\om > 0.31$ (see Figure 1). 
This supernovae/CMB inconsistency is strong motivation to use CMB data to 
explore a larger parameter space which includes $\ol$.
If the inconsistency is caused by the incorrect assumption that $\ol = 0$, then
such an analysis will show it.
The result of the analyis presented here is that $\ol > 0$ can resolve this inconsistency.

Testing large parameter spaces is important to minimize the model dependence of the results.
For example, in Lineweaver \& Barbosa (1998a) the CMB data favored $\h= 0.30^{+0.18}_{-0.07}$ 
({\bf if} $\om = 1$, {\bf if} $\ol = 0$ and {\bf if} all the other 
assumptions made are valid). In Lineweaver \& Barbosa (1998b), hereafter LB98b,
we dropped the $\om = 1$ assumption and still found low $\h$ values ($\h = 0.40$  but 
with large error bars: $0.26 < \h < 0.97$) and $\om > 0.57$.
Thus the  CMB data prefer high $\om$ values ({\bf if} $\ol=0$).
These may be big {\bf if} 's.
The purpose of this paper is to make these {\bf if} 's smaller by
exploring a still larger region of parameter space.

Other workers have used CMB observations to constrain cosmological 
parameters in CDM models ( e.g. Bond \& Jaffe 1997, deBernardis \etal 1997, 
Ratra \etal 1997, Hancock \etal 1998, Lesgourgues \etal 1998, Bartlett \etal 1998,
Webster \etal 1998, White 1998).
The previous work most similar to this Letter is White (1998).
White (1998) combined supernovae results with the Hancock \etal (1998) estimate of the
position in $\ell-$space of the peak in the CMB power spectrum. 
In Section 5, I compare my results to White (1998) and other work.

\section{Method}
\label{sec:method}

I use $\chi^{2}$ minimization to identify the best fit and 
$\Delta \chi^{2} = [1,4,9,16]$ to identify  $[68.3\%, 95.4\%, 99.7\%,99.9\%]$
confidence regions around the minima. The $\chi^{2}$ computation is
\be
\chi^{2}({\mathbf \theta}) = \sum_{i} 
\frac{
(model({\mathbf \theta})- data_{i})^{2}
}
{\sigma_{i}^{2}},
\ee
where ${\mathbf \theta}= [\ol, \om, \h, \obh, \n, \Q]$ which are, respectively, the
cosmological constant, the matter density, Hubble's constant, the baryonic density, 
the primordial power spectrum index and the primordial power spectrum amplitude 
at $\ell = 10$.
$\ol$ and $\om$ are normalized to the critical density ($\ol = \l/3 H_{o}^{2}$,
$\om = \rho_{m}/\rho_{crit}$) and $\h$ is the dimensionless
Hubble constant normalized at $100\: km\; s^{-1}\;Mpc^{-1}$.
The following ranges for these parameters were used:
$0 \le \ol \le 0.9,\: 0.1 \le \om \le 1.0,\: 0.15 \le \h \le 1.00,\:
 0.005 \le \obh \le 0.030,\: 0.58 \le \n  \le 1.42,
14.0\: \mu$K $\le \Q \le 23\; \mu$K 
with respective stepsizes 
$0.05, 0.05, 0.05, 0.005, 0.03, 0.5$. 
Although I am exploring a 6 dimensional parameter space
I limit the discussion in this {\it Letter} to the $\om - \ol$ plane.
A $\chi^{2}$ value is calculated for all points in the ranges above,  
consistent with the condition $\ob \le \om$. All 6 parameters are free to
take on any value which minimizes the $\chi^{2}$. 
See \S 2 of Lineweaver \etal (1997) and \S 2 of LB98b for more details of the method.
Compared to LB98b the improvements here are:
1) inclusion of $\ol$ as an extra dimension of parameter space
( not just flat$-\Lambda$),
2) new and updated CMB data points,
3) a larger range and higher resolution in the $\obh$ dimension,
4) the weak dependence of the helium fraction $Y_{He}$  on $\obh$ 
is included (Sarkar 1996, Hogan 1996).
Previously we had just set $Y_{He}= 0.24$ for all values of $\obh$.

The models in Eq. (1) were computed with
CMBFAST (Seljak \& Zaldarriaga 1996). The sum on $i$ in Eq. (1) is over
35 independent CMB anisotropy measurements. These data points
are given in Table 1 of LB98b with the following updates:
two new  points from Femen\'{\i}a \etal (1998), updated values from Baker (1998) and Leitch (1998)
to conform to the final published results and improved estimates of the error bars on the earlier
MSAM results. The previous 5 MAX points have been combined into one point (Tanaka 1997) and
I use the uncorrelated DMR points reported in Tegmark \& Hamilton (1997).
I also now include the data point from Picirillo \& Calisse (1993). 
In this analysis I use the Leitch (1998) recalibration of the Saskatoon 
measurements (Netterfield \etal 1997) which involves a $5\%$ increase in amplitude with
a $4\%$ dispersion about the new central values (see Section 2.3 of LB98b).

I have compared the results from LB98b with results from this updated data set
in open models ($\ol=0$).
The difference is small.
LB98b reported $ 0.26 < \h < 0.97$ and $\om > 0.53$, ($\om > 0.31$ at 99.9\% confidence level) 
with central values of $\h = 0.40$ and $\om = 0.85$.
The analogous result with the new data set is
$ 0.30 < \h < 0.98$ and  $\om > 0.57$ ($\om > 0.34$ at 99.9\% confidence level)
with central values $\h = 0.55$ and $\om = 0.75$.

\section{CMB Results}

The main result of the  CMB analysis is shown in Fig. 2.
The CMB data prefer the narrow hashed region (68.3\% confidence level) enclosed by 
approximate 95.4\%, 99.7\% and 99.9\% contours. 
The `X' marks the best fit: $(\om,\ol) = (0.45,0.35)$.
The minimum $\chi^{2}$ value is 22.1.
With normally distributed errors, the probability of obtaining a $\chi^{2}$ this low or lower
is $22.3\%$.

Since we have not considered closed models, the 68.3\% confidence region is cut off
by the $\om + \ol = 1$ limit. However the $\chi^{2}$ values along this
cutoff are very close to the 68.3\% confidence level.
Thus when the entire square is explored one should expect the 68.3\% confidence 
region to widen only for $\om \gsim 0.5$ and only by a narrow strip approximately 
parallel to the $\om + \ol = 1$ line.
If one restricts the analysis to flat-$\l$ models, the best-fit is
$\ol = 0.1$ with an upper limit $\ol < 0.61$.

This analysis also yields new constraints on the power spectral index and on the 
power spectrum normalization. At the  $\chi^{2}$ minimum  
$\n = 1.06^{+0.12}_{-0.18}$ and $\Q = 18 \pm 1 \; \mu$K.
The best-fit Hubble's constant is  $h = 0.75$ 
but with a large 68.4\% confidence range $0.35 < \h < 0.98$.
Thus $\h$ is not very usefully constrained by the CMB alone.
The $\chi^{2}$ minimum is obtained at the maximum value considered for the baryonic density:
$\obh = 0.030$ with a 68.4\% lower limit of $0.025$.
Thus the  CMB prefers high $\obh$ values.

\section{Combining Constraints}

Current observational constraints from supernovae, cluster mass-to-light ratios and
double radio sources in the $\om - \ol$ plane are given
in Table 1. To approximate the region of the $\om - \ol$ plane favored by 
non-CMB observations, I form likelihoods from the limits in Table 1 and from
published contours (e.g. Riess \etal 1998, Fig. 6, Carlberg \etal 1998, Fig. 1,
Daly \etal 1998, Fig. 1). I then form joint likelihoods  
${\mathcal L}_{non-CMB} = {\mathcal L}_{SN} \times {\mathcal L}_{Clusters} \times
{\mathcal L}_{radio}$, where all terms are functions of $\om$ and $\ol$.  
The CMB results are combined with the non-CMB results in the same way:
${\mathcal L}_{tot}= {\mathcal L}_{CMB} \times {\mathcal L}_{non-CMB}$.

There are a variety of ways in which the limits can be selected and combined.
My strategy is to be reasonably conservative by trying not to over-constrain parameter space.
Practically this means using contours large enough to include possible systematic errors. 
For example, two independent supernovae groups are in the process of taking data and 
refining their analysis and calibration techniques.
Table 1 lists their current limits (ref. 3 \& 6) which are consistent.
I combine each supernovae result separately with the CMB
constraints and list the result in the row marked `+ CMB' directly under the supernovae result.
However, the main result I quote in the abstract (last row of Table 1) comes from using 
the least-constraining of the two (ref. 3) in the conservative combination of non-CMB constraints,
i.e., ${\mathcal L}_{tot}= {\mathcal L}_{CMB} \times {\mathcal L}_{non-CMB(conservative)}$.
Figure 3 shows each of the three terms in this equation.
 
I apply the same strategy with the Carlberg \etal (1997, 1998) cluster mass-to-light
ratios. They report 30\% errors in their $\om$ result but also
cite a ``worst case'' of a 73\% error if all the systematic errors conspire and 
add linearly. In Table 1, I give the result of combining
the 30\% and 73\% versions separately with the CMB constraints. 
However, I use the 73\% error in the conservative
combination of non-CMB constraints. Thus the 
$\ol = 0.62 \pm 0.16$ and $\om = 0.24 \pm 0.10$ quoted in the abstract
are conservative in the sense that the error bars from the SNIa and
cluster mass-to-light ratios are ``worst case'' error bars.
A summary of the systematic error analysis of the cluster mass-to-light
ratios result is given in Section 9 of Carlberg \etal (1997)
and of the supernovae results in Section 5 of Riess \etal (1998).

The conservative result we quote is robust in the sense that when any one of 
these non-CMB constraints is combined with the CMB constraints,
similar results are obtained: $\ol = 0$
is more than the $95.4\%$ confidence level away from the best-fit.
Systematic errors may compromise one or the other of the 
observations but are less likely to bias all of the 
observations in the same way.

Those confident in the Riess \etal (1998) and the Carlberg \etal 
(1997, 1998) results should quote the extremely tight limits
labelled `optimistic' in the  penultimate row of Table 1:
$\om = 0.18 \pm 0.04$ and
$\ol = 0.71^{+0.07}_{-0.08}$.
In this small region of the $\om - \ol$ plane
the CMB data prefer
$\h = 0.80 \pm 0.10$, $\Q = 18 \pm 0.5\: \mu$K, $\n = 1.0 \pm 0.1$ and $\obh \sim  0.025$.
%

{\scriptsize
\begin{table}
\begin{center}
\caption{ Non-CMB and Non-CMB + CMB Constraints on $\ol$ and $\om$}
\begin{tabular}{|l|l|l|l|l|l|} \hline
\multicolumn{1}{|l}{Reference) Method } &
\multicolumn{1}{|c}{$\ol = 0$ } &
\multicolumn{2}{|c|}{$      \om+\ol=1            $} &
\multicolumn{2}{|c|}{$      \om+\ol \le 1        $}\\
\tableline
\multicolumn{1}{|c}{ }&
\multicolumn{1}{|c}{$\om$} &
\multicolumn{1}{|c}{$\om$} &
\multicolumn{1}{|c}{$\ol$} &
\multicolumn{1}{|c}{$\om$} &
\multicolumn{1}{|c|}{$\ol$}\\
\tableline
1) SNIa                          &$ 0.88^{+0.69}_{-0.60}$&$ 0.94^{+0.34}_{-0.28}      $&$ 0.06^{+0.28}_{-0.34}$&$                     $&$                     $\\
2) SNIa                          &$-0.2  \pm 0.4        $&$ 0.6  \pm 0.2              $&$ 0.4 \pm 0.2         $&$                     $&$                     $\\
3) SNIa                          &$-0.4  \pm 0.51^{a}   $&$ 0.27 \pm 0.3^{b}          $&$ 0.73 \pm 0.3        $&$                     $&$                     $\\
3) SNIa + CMB                    &$                     $&$                           $&$                     $&$ 0.33^{+0.15}_{-0.18}$&$ 0.52^{+0.25}_{-0.22}$\\

\hline
4) SNIa                          &$-0.1  \pm 0.5        $&$ 0.65 \pm 0.3^{c}          $&$ 0.35 \pm 0.3        $&$                     $&$                     $\\
5) SNIa                          &$-0.2^{+1.0}_{-0.8}   $&$ 0.4^{+0.5}_{-0.4}         $&$ 0.6^{+0.4}_{-0.5}   $&$                     $&$                     $\\
6) SNIa                          &$-0.35 \pm 0.18       $&$ 0.24^{+0.17}_{-0.10}{}^{d}$&$ 0.76^{+0.10}_{-0.17}$&$                     $&$                     $\\
6) SNIa + CMB                    &$                     $&$                           $&$                     $&$0.26^{+0.09}_{-0.11} $&$ 0.63^{+0.14}_{-0.15}$\\

\hline
7) Cluster M/L                   &$ 0.19 \pm  0.14^{e}  $&$                           $&$                     $&$                     $&$                     $\\
7) Cluster M/L + CMB             &$                     $&$                           $&$                     $&$0.24 \pm 0.10        $&$ 0.62^{+0.17}_{-0.19}$\\

7) Cluster M/L                   &$ 0.19 \pm  0.06^{f}  $&$                           $&$                     $&$                     $&$                     $\\
7) Cluster M/L + CMB             &$                     $&$                           $&$                     $&$0.17^{+0.04}_{-0.03} $&$ 0.73^{+0.07}_{-0.08}$\\

\hline
8) Double Radio Sources           &$-0.1^{+0.5}_{-0.4}   $&$     0.2^{+0.3}_{-0.2}      $&$  0.8^{+0.2}_{-0.3}  $&$                     $&$                     $\\
8) Double Radio Sources + CMB     &$                     $&$                           $&$                     $&$0.33^{+0.16}_{-0.18}  $&$ 0.52 \pm 0.25       $\\

\hline
Non-CMB(optimistic)$^{g}        $&$                     $&$                           $&$                     $&$0.16 \pm 0.05         $&$ 0.57 \pm 0.25      $\\
Non-CMB(conservative)$^{h}      $&$                     $&$                           $&$                     $&$0.15 \pm 0.13         $&$ 0.46^{+0.41}_{-0.41}$\\

Non-CMB(optimistic) + CMB$^{g}  $&$                     $&$                           $&$                     $&$0.18 \pm 0.04         $&$ 0.71^{+0.07}_{-0.08}$\\
Non-CMB(conservative) + CMB$^{h}$&$                     $&$                           $&$                     $&${\mathbf 0.24 \pm 0.10} $&${\mathbf 0.62 \pm 0.16}$\\
\tableline 
\end{tabular}\\
\end{center}

\tiny
\noindent
$^{a}$ $-0.4 \pm 0.1 \pm 0.5$  (statistical and systematic errors respectively) I have added the statistical and systematic errors in quadrature\\
$^{b}$ $0.27 \pm 0.06 \pm 0.3$ (statistical and systematic errors respectively) I have added the statistical and systematic errors in quadrature\\
$^{c}$ as quoted in Riess \etal 1998\\
$^{d}$ Riess \etal (1998), Figure 6 (`MLCS method' + `snapshot method') 
using either the solid or dotted contours whichever is larger (corresponding respectively to the analysis
with and without SN1997ck).\\
$^{e}$ ``worst case'' result with $73\%$ errors (Carlberg \etal 1997, p 473) \\
$^{f}$  30\% error cited as the main result\\
$^{g}$ optimistic combined constraints using SNIa results from ref. 6) (rather than ref. 3) and using
the 30\% error bars (rather than the 73\% error bars) from Carlberg \etal (1997)\\
$^{h}$ conservative combined constraints using SNIa results from ref. 3) (rather than ref. 6) and using
the 73\% error bars (rather than the 30\% error bars) from Carlberg \etal (1997)\\

1) Perlmutter \etal (1997),
2) Perlmutter \etal (1998),
3) Perlmutter  private communication (1998),
4) Garnavich \etal (1998),
5) Schmidt \etal  (1998),
6) Riess \etal (1998),  
7) Carlberg \etal (1997) and Carlberg \etal (1998),
8) Daly \etal (1998) 

\end{table}
}   
\normalsize

It would be useful to add constraints on $\ol$ from lensing. 
Although Kochanek (1996) and Falco, Kochanek \& Mu\~{n}oz (1998)
report $\ol \lsim 0.7$, the lensing estimates from
Chiba \& Yoshii (1997) ($\ol \sim 0.8$) and Fort \etal (1997) 
($\ol \sim 0.6$) are in agreement with the result found here.
Thus lensing limits on $\ol$ still seem too uncertain to add 
much to the analysis.
However, when I fold $\ol < 0.7$ into the analysis the result for $\ol$ 
comes down less than $0.5\; \sigma$.

\section{Comparison with Previous Results}

White (1998) combined his supernovae analysis  with CMB results and pointed out the
important complementarity of the two. He used the Hancock \etal (1998) estimate of the
position in $\ell-$space of the peak in the CMB power spectrum based on 
$\Omega_{o}(=\om + \ol)-$dependent (but not $\ol-$dependent) phenomenological models introduced by
Scott, Silk \& White (1995). The Hancock \etal (1998) result is: $\lp = 263^{+139}_{-94}$.
This should be compared to the LB98b result of $\lp = 260^{+30}_{-20}$.
The tighter limits are presumably due to the more precise parameter dependencies of the power
spectrum models and the more recent data set.
In Figure 3 of White (1998), the $1\sigma$ contours from supernovae and CMB
overlap in a region consistent with the results reported here.

Another result of the Hancock \etal (1998) analysis is
$\om + \ol = 0.7^{+0.8}_{-0.5}$. This is consistent with
the CMB results presented here. A rough approximation of the elongated
triangle in Figure 2 is a strip parallel to the $\om + \ol = 1$ line
approximated by $\om + \ol = 0.8 \pm 0.3$.

By combining CMB and IRAS power spectral constraints,
Webster \etal (1998) obtain $\om = 0.32 \pm 0.08$ in flat-$\l$ models with $\n = 1$.
This is in very good agreement with the values obtained here from the combination
of CMB, supernovae, cluster mass-to-light ratios and double radio sources.

The result presented in this {\it Letter} is consistent with a large and respectable subset of 
observational constraints (Turner 1991, Ostriker \& Steinhardt 1995,
Roos \& Harun-or-Rachid 1998).
However models in this region of the $\om - \ol$ plane appear to have problems 
fitting the shape of the large-scale power spectrum measured from the APM 
survey around $k = 0.1\: \h\: Mpc^{-1}$ (Peacock 1998)
and are in disagreement with constraints on $\om$ from the POTENT analysis
of the local velocity field (Dekel \etal 1997).

We have assumed Gaussian adiabatic fluctuations. However 
Ferreira \etal (1998)  have analyzed the DMR four-year maps
and found tentative evidence for non-Gaussianity.
Peebles (1998) has presented  the case for isocurvature
rather than adiabatic initial conditions.
If either non-Gaussian processes or isocurvature initial conditions
play significant roles in CMB anisotropy formation, then the CMB results presented
here are significantly compromised.


\section{Summary}

The results presented here are largely observational but are model dependent.
In a series of papers (Lineweaver \etal 1997, Lineweaver \& Barbosa 1998a, 1998b), and
now in this work, we have looked at increasingly larger regions of parameter space.
Each time the $\chi^{2}$ minimum has been found to lie within the new region.
This might be taken as a sign of caution not to take the currently favored region
too seriously. On the other hand, our choice of new parameter space to explore
has been guided by independent observational results.

I have used the most recent CMB data to constrain the leading 
CDM models in the $\om - \ol$ plane.
A narrow triangular region is preferred.
This triangle is elongated in such a way that its intersection
with even conservative versions of other constraints
is small and provides the current best limits in the plane: 
$\ol = 0.62 \pm 0.16$ and $\om = 0.24 \pm 0.10$. 
This complementarity between CMB and other observations rules out $\ol = 0$ 
models at more than the $99\%$ confidence level.

Until recently observations could not discriminate between a zero and 
a non-zero cosmological constant.
However a wide range of observations have indicated that $\om < 1$ and 
the most recent observations appear to favour $\ol > 0$.
The addition of the CMB constraints presented here to these other cosmological observations
strengthens this conclusion substantially. 
 
I gratefully acknowledge discussions with 
Saul Perlmutter and Brian Schmidt about the supernovae data sets and
I acknowledge Uros Seljak and Matias Zaldarriaga for providing the 
Boltzmann code. I am supported by a Vice-Chancellor's fellowship 
at the University of New South Wales, Sydney, Australia.

\clearpage
{\normalsize    

}

\clearpage
\begin{figure*}[!b]
\centerline{\psfig{figure=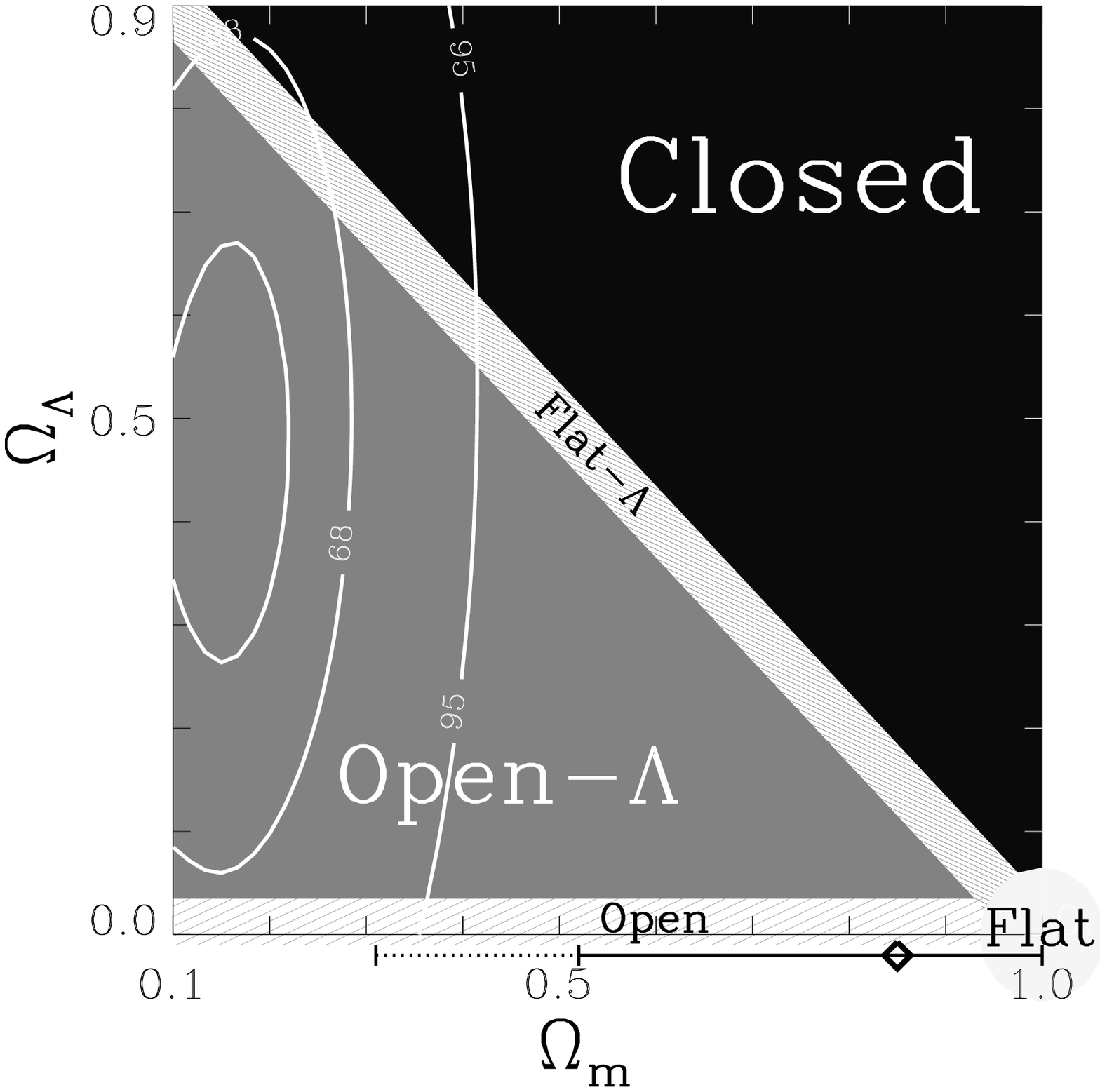,height=16.0cm,width=16.cm}}
\caption{Overview of cold dark matter (CDM) models in the $\om - \ol$ plane.
This parameter space includes the popular cosmological candidates: flat, 
flat-$\Lambda$ and open models as well as the less popular open$-\Lambda$ models. 
Closed models (upper right triangle) are not considered here.
The white contours indicate the region favored by the joint likelihood of constraints from  
recent supernovae,
cluster mass-to-light ratios,
and double radio sources.
They are approximate $38.3\%, 68.3\%$ and $95.4\%$ confidence regions (see Table 1 and Section 4).
In open models ($\ol = 0$) there is a significant inconsistency between CMB results and these
other observations.
Previously reported constraints from the CMB in open models are indicated just below the 
$\om$ axis: $\om = 0.85$ with a $68.3\% $ lower limit of $0.53$ and 
a $99.9\%$ lower limit of $0.31$ (Lineweaver \& Barbosa 1998b).}
\label{fig:figone}
\end{figure*}
\clearpage
\begin{figure*}[!b]
\centerline{\psfig{figure=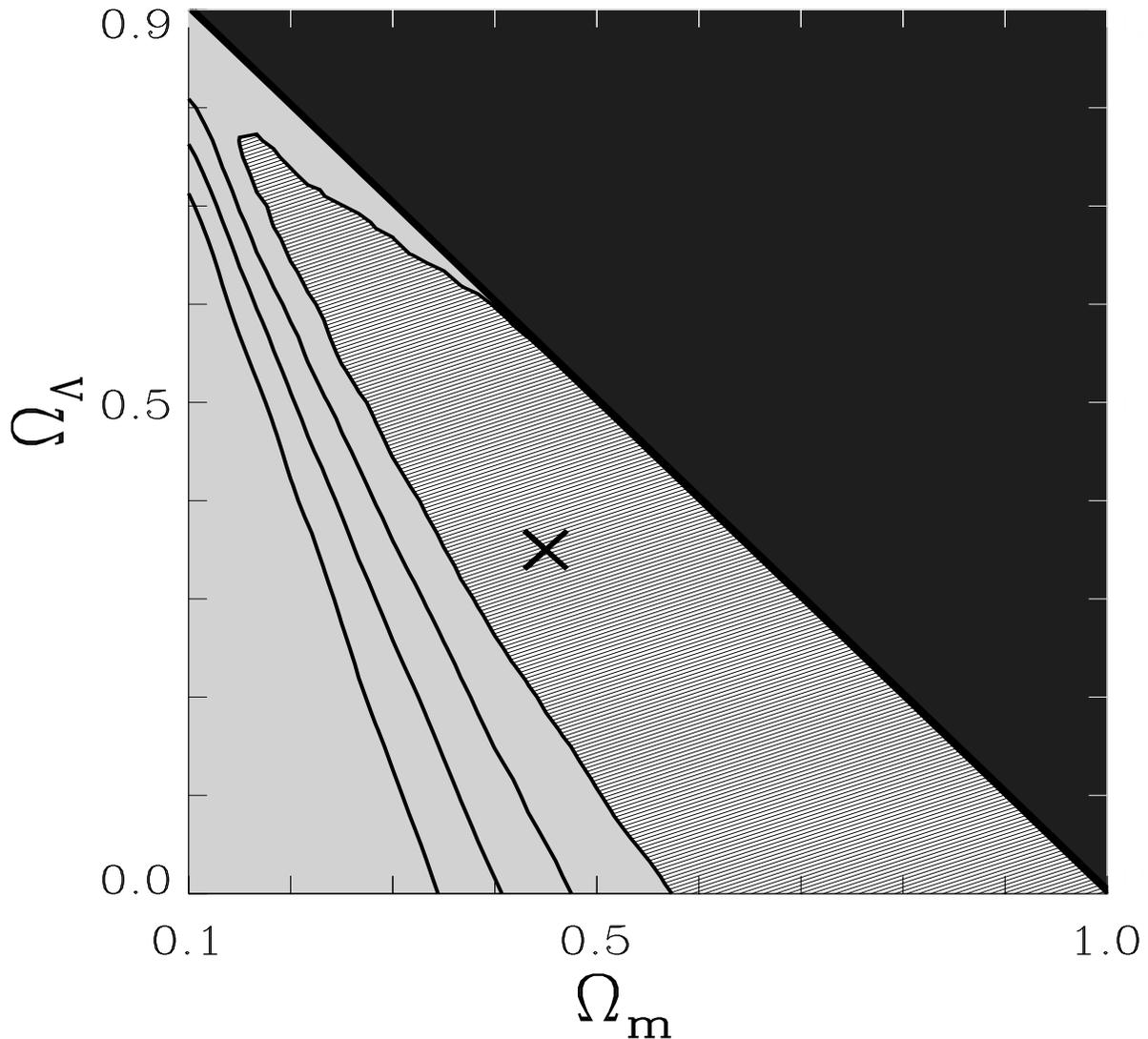,height=16.0cm,width=16.cm}}
\caption{
The current CMB data prefer the narrow hashed triangular region (68.3\% confidence level) 
enclosed by approximate $95.4\%, 99.7\%$ and $99.9\%$ contours. 
The `X' marks the best fit: $(\om,\ol) = (0.45,0.35)$.
This region is bounded by the limits: $\ol < 0.77$ and $\om > 0.15$. Assuming $\ol = 0$, 
yields the lower limit $\om > 0.57$. Restricting consideration to flat-$\l$ models,
the best-fit is $\ol = 0.1$ with an upper limit $\ol < 0.61$.
}
\label{fig:figtwo}
\end{figure*}
\clearpage
\begin{figure*}[!b]
\centerline{\psfig{figure=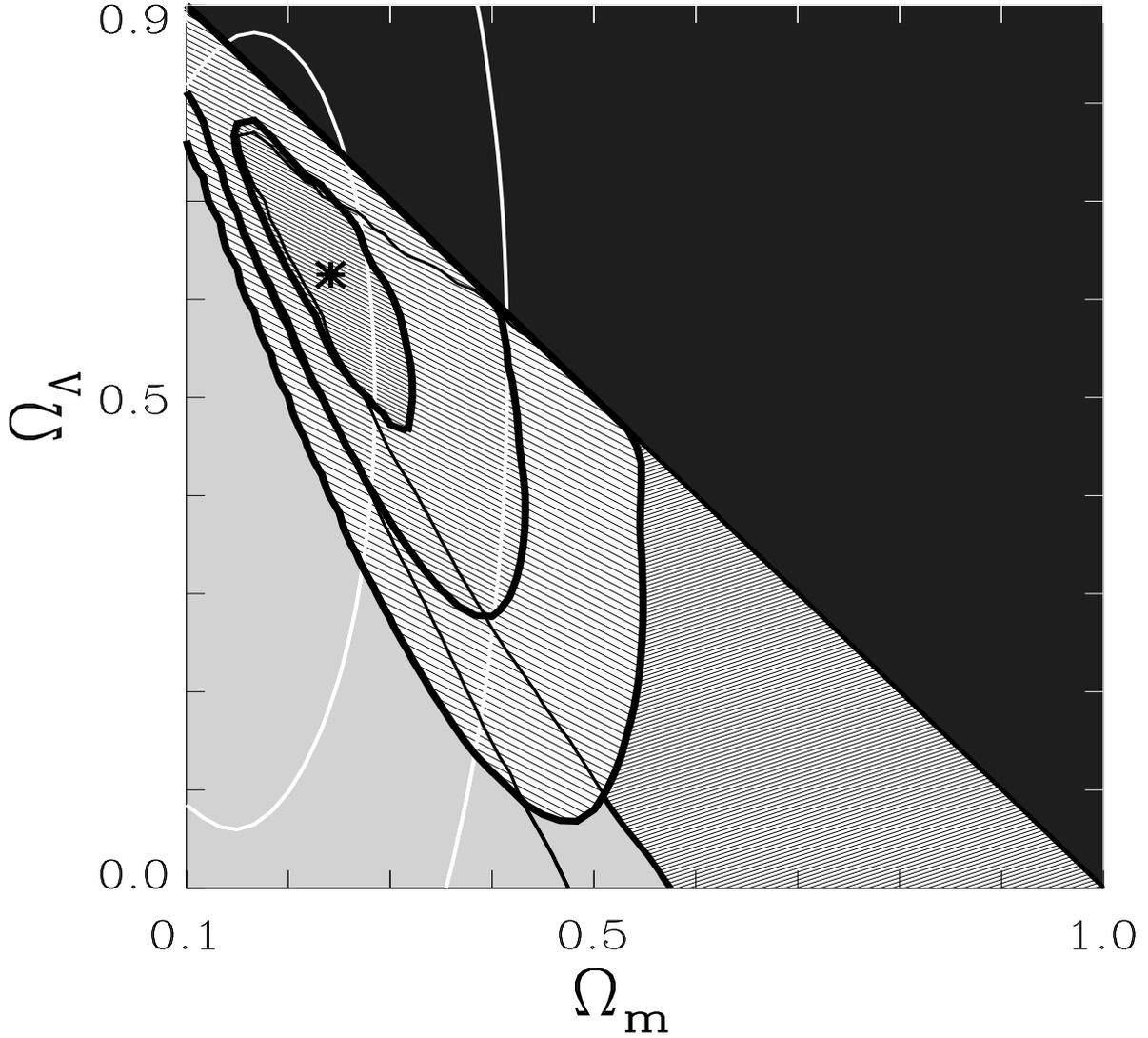,height=16.0cm,width=16.cm}}
\caption{
The thick dark lines are the approximate $68.3\%$, $95.4\%$ and $99.7\%$ contours from the
joint likelihood of the CMB and non-CMB constraints.
The $68.3\%$ and $95.4\%$ confidence levels from non-CMB observations are
in white (same as in Figure 1).
The $68.3\%$ and $95.4\%$ confidence levels from CMB observations are
the thin black lines (same as in Figure 2 but partially obscured here).
The best fit is : $\om = 0.24 \pm 0.10$ and $\ol = 0.62 \pm 0.16$.
$\ol = 0$ models are excluded at more than the $99.7\%$ confidence level.
}
\label{fig:figthree}
\end{figure*}

\end{document}